\documentclass[prb,aps]{revtex4}

\usepackage[dvips]{graphics}
\usepackage{amssymb}
\usepackage{psfig}

\DeclareGraphicsExtensions{.bmp,.eps,.gif,.png}

\begin{document}

\title{Spin-polarized currents in superconducting films}
\author{M. Bo\v{z}ovi\'{c}\footnote{{\em E-mail}: {\tt mbozovic@infosky.net}} and Z. Radovi\'c\footnote{{\em E-mail}: {\tt zradovic@ff.bg.ac.yu}}}

\address{Department of Physics, University of Belgrade, P.O. Box 368, 11001 Belgrade, Yugoslavia}


\begin{abstract}
We present a microscopic theory of coherent quantum transport
through a superconducting film between two ferromagnetic
electrodes. The scattering problem is solved for the general case
of ferromagnet/superconductor/ferro\-magnet (FSF) double-barrier
junction, including the interface transparency from metallic to
tunnel limit, and the Fermi velocity mismatch. Charge and spin
conductance spectra of FSF junctions are calculated for parallel
(P) and antiparallel (AP) alignment of the electrode
magnetization. Limiting cases of nonmagnetic normal-metal
electrodes (NSN) and of incoherent transport are also presented.
We focus on two characteristic features of finite size and
coherency: subgap tunneling of electrons, and oscillations of the
differential conductance. Periodic vanishing of the Andreev
reflection at the energies of geometrical resonances above the
superconducting gap is a striking consequence of the quasiparticle
interference. Also, the non-trivial spin-polarization of the
current is found for FSF junctions in AP alignment. This is in
contrast with the incoherent transport, where the unpolarized
current is accompanied by excess spin accumulation and destruction
of superconductivity. Application to spectroscopic measurements of
the superconducting gap and the Fermi velocity is also discussed.
\end{abstract}

\maketitle


\section{Introduction}

During the past decade, there has been a growing interest in
various electronic systems driven out of equilibrium by the
injection of spin-polarized carriers. Such systems can be realized
by current-biasing structures consisting of ferromagnetic and
non-ferromagnetic (e.g. superconducting) layers, due to the
difference in population of majority and minority spin
subbands.\cite{Prinz} The concept of spin-polarized current
nowadays has attracted considerable interest in ferromagnetic
heterostructures, in particular for applications in
spintronics.\cite{Osofsky}

Charge transport through a normal metal/superconductor (NS)
junction, with an insulating barrier of arbitrary strength at the
interface, has been studied by Blonder, Tinkham, and Klapwijk
(BTK),\cite{BTK} and the Andreev reflection is recognized as the
mechanism of normal-to-supercurrent
conversion.\cite{Andreev,Furusaki Tsukada} The BTK theory has been
extended by Tanaka and Kashiwaya to include the anisotropy of the
pair potential in $d$-wave superconductors.\cite{Tanaka 95,Tanaka
00} The modification of the Andreev reflection by the spin
injection from a ferromagnetic metal into a superconductor in
ferromagnet/superconductor (FS) junctions was first analyzed by de
Jong and Beenakker.\cite{dJB} More recently, the effects of
unconventional $d$-wave and $p$-wave pairing and of the exchange
interaction in FS systems, such as the zero-bias conductance peak
and the virtual Andreev reflection, have been clarified by
Kashiwaya {\it et al}.\cite{Beasley} and Yoshida {\it et
al}.\cite{Yoshida} The Fermi velocity mismatch between two metals
can also significantly affect the Andreev reflection by altering
the subgap conductance,\cite{Zutic} which is similar to the
presence of an insulating barrier.\cite{Zhu}

In experiments, a superconductor is used to determine the spin
polarization of the current injected from (or into) a ferromagnet
by measuring the differential conductance. These measurements have
been performed on tunnel junctions in an external magnetic
field,\cite{Tedrow,Platt} metallic point
contacts,\cite{Soulen,Novo} nano-contacts formed by
microlithography,\cite{Upad} and FS junctions with $d$-wave
superconductors, grown by molecular beam epitaxy.\cite{Vasko} In
diffusive FS junctions, the excess resistance may be induced by
spin accumulation near the insulating interface,\cite{Jedema} and
by the proximity effect.\cite{Gueron,Petrashov,Sillanpaa}

When electrons pass incoherently through the interfaces, the BTK
model can be successfully applied to normal
metal/superconductor/normal metal (NSN) or
ferromagnet/superconductor/ferromagnet (FSF) double
junctions.\cite{Takahashi,Kinezi} However, the properties of
coherent quantum transport in clean superconducting
heterostructures are strongly influenced by size effects, which
are not included in the BTK model. Well-known examples are the
current-carrying Andreev bound states\cite{Tanaka 00,Nazarov} and
multiple Andreev reflections\cite{KBT,Basel,Ingerman} in
superconductor/normal metal/super\-conductor (SNS) junctions.
Since early experiments by Tomasch,\cite{Tomasch} the geometric
resonance nature of the differential conductance oscillations in
SNS and NSN tunnel junctions has been ascribed to the electron
interference in the central
film.\cite{Anderson,Rowell,McMillan,Kanadjani} Recently, the
McMillan-Rowell oscillations were observed in SNS edge junctions
of $d$-wave superconductors, and used for measurements of the
superconducting gap and the Fermi velocity.\cite{Nesher}

Here we present a comperhensive microscopic theory of coherent
transport in FSF double junctions (with NSN as a special
case).\cite{Milos,ZZM} We limit ourselves to clean conventional
($s$-wave) superconductors, and neglect, for simplicity, the
self-consistency of the pair potential,\cite{Geers,Buzdin} and
nonequlibrium effects of charge and spin accumulation at the
interfaces.\cite{FalkoB,McCann} When two interfaces are recognized
by electrons simultaneously, characteristic features of finite
size and coherency are the subgap tunneling of electrons and
oscillations of both charge and spin differential conductances
above the gap. One consequence of the quasiparticle interference
is the periodic vanishing of the Andreev reflection at the
energies of geometrical resonances. The other is the existence of
a non-trivial spin-polarization of the current not only for the
parallel (P), but also for the antiparallel (AP) alignment of the
electrode magnetizations. Previous analysis of incoherent
transport in FSF double junctions predict the absence of the spin
current and suppression of superconductivity with increasing
voltage for AP alignment, as a result of spin imbalance in the
superconducting film.\cite{Takahashi,Kinezi}

\bigskip

\section{The scattering problem}

\begin{figure}[b]
\dimen255=0.7\textwidth
    \centerline{\psfig{figure=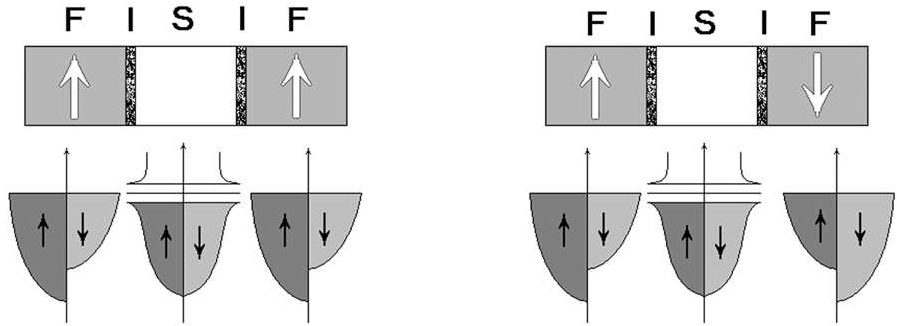,height=0.4\dimen255}}
    \caption{Double barrier junction consisting of two ferromagnets (F) and a
            superconductor (S) separated by insulating barriers (I), for
            parallel (left panel) and antiparallel (right panel) alignment of
            the electrode magnetization. Schematic of the corresponding
            densities of states.}
    \label{sch}
\end{figure}

We consider an FSF double junction consisting of a clean
superconducting layer of thickness $l$, connected to ferromagnetic
electrodes by thin, insulating interfaces, Fig. \ref{sch}. For the
ferromagnetic metal we adopt the Stoner model describing the
spin-polarization effect by the usual one-electron Hamiltonian
with an exchange potential. The quasiparticle propagation is
described by the Bogoliubov--de Gennes equation
\begin{eqnarray}
\left(
\begin{array}{ccc}
  H_0({\bf r})-\rho_{\sigma}h({\bf r}) && \Delta({\bf r}) \\
  \Delta^{*}({\bf r}) && -H_0({\bf r})+\rho_{\bar{\sigma}}h({\bf
r})
\end{array}
\right) \Psi_\sigma({\bf r})~=~E\Psi_\sigma({\bf r}), \label{BdG}
\end{eqnarray}
with $H_{0}({\bf r})=-\hbar^{2}\nabla^{2}/2m+W({\bf r})+U({\bf
r})-\mu$, where $U({\bf r})$ and $\mu$ are the Hartree and the
chemical potential, respectively. The interface potential is
modeled by $W({\bf r})=\hat{W}\{\delta(z)+\delta(z-l)\}$, where
$z$-axis is perpendicular to the layers and $\delta(z)$ is the
Dirac $\delta$-function. Neglecting the self-consistency of the
superconducting pair potential, $\Delta({\bf r})$ is taken in the
form $\Delta \Theta(z) \Theta(l-z)$, where $\Delta$ is the bulk
superconducting gap and $\Theta(z)$ is the Heaviside step
function. In Eq. (\ref{BdG}), $\sigma$ is the quasiparticle spin
($\sigma =\uparrow ,\downarrow$~and
~$\bar{\sigma}=\downarrow,\uparrow$), $E$ is the energy with
respect to $\mu$, $h({\bf r})$ is the exchange potential given by
$h_{0}\{\Theta(-z)+[-]\Theta(z-l)\}$ for the P [AP] alignment, and
$\rho_{\sigma}$ is 1 (-1) for  spins up (down). The electron
effective mass $m$ is assumed to be the same for the whole
junction. Here, $\mu-U({\bf r})$ is the Fermi energy of the
superconductor, $E^{(S)}_F$, or the mean Fermi energy of a
ferromagnet, $E^{(F)}_F=(E^\uparrow_F+E^\downarrow_F)/2$. Moduli
of the Fermi wave vectors, $k^{(F)}_F= \sqrt{2mE^{(F)}_F/\hbar^2}$
and $k^{(S)}_F= \sqrt{2mE^{(S)}_F/\hbar^2}$, can be different in
general, and in the following, the Fermi wave vector mismatch
(FWVM) will be taken into account through the parameter
$\kappa=k^{(F)}_F/k^{(S)}_F$. The parallel component of the wave
vector ${\bf k}_{||,\sigma}$ is conserved, and the wave function
\begin{equation}
\Psi_\sigma({\bf r})=\exp(i{\bf k}_{||,\sigma} \cdot {\bf r}
  )~\psi_\sigma(z),
\end{equation}
satisfies the boundary conditions
\begin{eqnarray}
\label{bc1}
\psi_\sigma (z)|_{z=0_-}&=&\psi_\sigma (z)|_{z=0_+},\\
\frac{d\psi_\sigma (z)}{dz}\Big|_{z=0_-}&=&\frac{d\psi_\sigma
(z)}{dz}\Big|_{z=0_+} -\frac{2m\hat{W}}{\hbar ^2}\psi_\sigma
(0),\\
\psi_\sigma (z)|_{z=l_-}&=&\psi_\sigma (z)|_{z=l_+},\\
\frac{d\psi_\sigma (z)}{dz}\Big|_{z=l_-}&=&\frac{d\psi_\sigma
(z)}{dz}\Big|_{z=l_+}-\frac{2m\hat{W}}{\hbar ^2}\psi_\sigma (l)
\label{bc4}.
\end{eqnarray}
Four independent solutions of Eq. (\ref{BdG}) correspond to the
four types of injection: an electron or a hole from either the
left or from the right electrode.\cite{Furusaki Tsukada}

For the injection of an electron from the left, with energy $E>0$,
spin $\sigma$, and angle of incidence $\theta$ (measured from the
$z$-axis), solution for $\psi_\sigma (z)$ in various regions has
the following form:

\noindent in the left ferromagnet ($z<0$)
\begin{equation} \psi_\sigma
(z)=\{\exp(ik^+_{\sigma}
z)+b_{\sigma}(E,\theta)\exp(-ik^+_{\sigma}
z)\}\left(\begin{array}{c}
     1 \\
     0 \\
   \end{array}\right)+
    a_{\sigma}(E,\theta)\exp(ik^-_{\bar{\sigma}} z)\left(\begin{array}{c}
      0 \\
      1 \\
    \end{array}\right),
\label{psiL}
\end{equation}
in the superconductor ($0<z<l$)
\begin{eqnarray}
\psi_\sigma (z)&=&\{
c_{1}(E,\theta)\exp(iq^+_{\sigma}z)+c_{2}(E,\theta)\exp(-iq^+_{\sigma}z)\}
\left(\begin{array}{c}
    \bar{u} \\
    \bar{v}
  \end{array}\right) \nonumber \\
&~&+\{
c_{3}(E,\theta)\exp(iq^-_{\sigma}z)+c_{4}(E,\theta)\exp(-iq^-_{\sigma}z)\}
\left(\begin{array}{c}
    \bar{v}^{*} \\
    \bar{u}^{*}
  \end{array}\right),
\label{psiS}
\end{eqnarray}
and in the right ferromagnet ($z>l$), for the P [AP] alignment of
the magnetizations
\begin{equation}
 \psi_\sigma(z)=c_{\sigma}(E,\theta)\exp(ik^+_{\sigma[{\bar\sigma}]}z)\left(\begin{array}{c}
     1 \\
     0 \\
   \end{array}\right)+
    d_{\sigma}(E,\theta)\exp(-ik^-_{{\bar\sigma}[\sigma]}z)\left(\begin{array}{c}
      0 \\
      1 \\
    \end{array}\right).
\label{psiR}
\end{equation}

Here, $\bar{u}=\sqrt{(1+\Omega/E)/2}$ and
$\bar{v}=\sqrt{(1-\Omega/E)/2}$ are the BCS coherence factors, and
$\Omega=\sqrt{E^2-\Delta^2}$. The $z$-components of the wave
vectors are $k^\pm_{\sigma}=\sqrt{(2m/\hbar
^2)(E^{(F)}_F+\rho_{\sigma}h_0 \pm E)-{\bf k}^2_{||,\sigma}}$, and
$q^\pm_{\sigma}=$ \\ $\sqrt{(2m/\hbar ^2)(E^{(S)}_F\pm\Omega)-{\bf
k}^2_{||,\sigma}}$, where $|{\bf k}_{||,\sigma}|=\sqrt{(2m/\hbar
^2)(E^{(F)}_F+\rho_{\sigma}h_0+E)}~\sin\theta$. The coefficients
$a_{\sigma}$, $b_{\sigma}$, $c_{\sigma}$, and $d_{\sigma}$ are,
respectively, the probability amplitudes of: (1) Andreev
reflection as a hole of the opposite spin (AR); (2) normal
reflection as an electron (NR); (3) transmission to the right
electrode as an electron (TE); (4) transmission to the right
electrode as a hole of the opposite spin (TH). Processes (1) and
(4) are equivalent to the formation of a Cooper pair in the
superconductor by taking one more electron from either the left or
the right electrode, respectively. Amplitudes of the Bogoliubov
electron-like and hole-like quasiparticles, propagating in the
superconducting layer, are given by the coefficients $c_1$ through
$c_4$.

From the probability current conservation, the probabilities of
outgoing particles satisfy the normalization condition
\begin{equation}
\label{ABCD}
A_\sigma(E,\theta)+B_\sigma(E,\theta)+C_\sigma(E,\theta)+D_\sigma(E,\theta)=1,
\end{equation}
where,
\begin{eqnarray}
\label{maliA}
A_\sigma(E,\theta)&=&\Re\left(\frac{\tilde{k}_{\bar{\sigma}}}
{\tilde{k}_{\sigma}}\right)|a_\sigma (E,\theta)|^2, \\
\label{maliB} B_\sigma(E,\theta)&=&|b_\sigma (E,\theta)|^2, \\
\label{maliC}
C_\sigma(E,\theta)&=&\Re\left(\frac{\tilde{k}_{\sigma[{\bar{\sigma}}]}}{\tilde{k}_{\sigma}}\right)
|c_\sigma(E,\theta)|^2, \\ \label{maliD} D_\sigma(E,\theta)&=&
\Re\left(\frac{\tilde{k}_{{\bar{\sigma}}[\sigma]}}{\tilde{k}_{\sigma}}\right)
|d_\sigma(E,\theta)|^2.
\end{eqnarray}

Neglecting small terms $E/E^{(F)}_F\ll 1$ and $\Delta/E^{(S)}_F\ll
1$ in the wave vectors, except in the exponents
\begin{equation}
\label{zeta} \zeta_\pm =l\left(q^+_{\sigma}\pm
q^-_{\sigma}\right),
\end{equation}
solutions of Eqs. (\ref{bc1})-(\ref{bc4}) for the probability
amplitudes can be written in the following form
\begin{eqnarray}
\label{a general} a_\sigma(E,\theta)&=&\frac{4
(\tilde{k}_{\sigma}/\tilde{q}_\sigma) \Delta
\sin(\zeta_-/2)}{\Gamma}\left[{\cal A}^R_+ E \sin(\zeta_-/2)+i
{\cal B}^R_+ \Omega \cos(\zeta_-/2)\right], \\ \label{b general}
b_\sigma(E,\theta)&=&\frac{1}{\Gamma}[{\cal A}^R_+{\cal
C}_+\Delta^2 - \left({\cal A}^R_+{\cal C}_+E^2 + {\cal B}^R_+{\cal
D}_+\Omega^2 \right)\cos(\zeta_-) + \left({\cal A}^R_-{\cal C}_- +
{\cal B}^R_-{\cal D}_-\right)\Omega^2\cos(\zeta_+) \nonumber\\ &~&
+ i\left({\cal B}^R_+{\cal C}_+ + {\cal A}^R_+{\cal
D}_+\right)E\Omega\sin(\zeta_-) - i\left({\cal B}^R_-{\cal C}_- +
{\cal A}^R_-{\cal D}_-\right)\Omega^2\sin(\zeta_+)], \\ \label{c
general} c_\sigma(E,\theta)&=&\frac{4
(\tilde{k}_{\sigma}/\tilde{q}_\sigma) \Omega
e^{-ik^+_{\bar{\sigma}}l } }{\Gamma}\times \nonumber\\
&~&\times\{ i\left[{\cal F}_+ \cos(\zeta_+/2)+i {\cal E}_+
\sin(\zeta_+/2)\right]E\sin(\zeta_-/2) - \left[{\cal
E}_+\cos(\zeta_+/2)+i {\cal F}_+\sin(\zeta_+/2)
\right]\Omega\cos(\zeta_-/2)\},
\\ \label{d general} d_\sigma(E,\theta)&=&\frac{4
(\tilde{k}_{\sigma}/\tilde{q}_\sigma) \Delta\Omega
e^{ik^-_{\sigma}l}}{\Gamma}\times \nonumber\\ &~&\times
i\left[{\cal F}_- \cos(\zeta_+/2)+i {\cal
E}_-\sin(\zeta_+/2)\right]\sin(\zeta_-/2),
\end{eqnarray}
where
\begin{eqnarray}
\label{Gamma} {\Gamma}={\cal A}^L_+{\cal A}^R_+\Delta^2 -
\left({\cal A}^L_+{\cal A}^R_+E^2 + {\cal B}^L_+{\cal
B}^R_+\Omega^2 \right)\cos(\zeta_-) + \left({\cal A}^L_-{\cal
A}^R_- + {\cal B}^L_-{\cal B}^R_-\right)\Omega^2\cos(\zeta_+)
\nonumber\\ + i\left({\cal A}^L_+{\cal B}^R_+ + {\cal B}^L_+{\cal
A}^R_+\right)E\Omega\sin(\zeta_-) - i\left({\cal A}^L_-{\cal
B}^R_- + {\cal B}^L_-{\cal A}^R_-\right)\Omega^2\sin(\zeta_+).
\end{eqnarray}
In Eqs. (\ref{a general})-(\ref{Gamma})
\begin{eqnarray*}
{\cal A}^{L(R)}_{\pm}=K^{L(R)}_1 \pm K^{L(R)}_2,&~~~ {\cal
B}^{L(R)}_{\pm}=1 \pm  K^{L(R)}_1 K^{L(R)}_2,&~~~ {\cal C}_{\pm}=
{K^{L}_1}^* \mp K^{L}_2, \\ {\cal D}_{\pm}=-(1 \mp {K^{L}_1}^*
K^{L}_2),&~~~ {\cal E}_{\pm}=K^{L}_2 \pm K^{R}_2,&~~~ {\cal
F}_{\pm}= 1 \pm K^{L}_2 K^{R}_2,
\end{eqnarray*}
with $K^L_1=\left(\tilde{k}_{\sigma}+iZ\right)/\tilde{q}_\sigma$,
$K^L_2=\left(\tilde{k}_{\bar{\sigma}}-iZ\right)/\tilde{q}_\sigma$,
$K^R_1=\left(\tilde{k}_{\sigma[{\bar{\sigma}}]}+iZ\right)/\tilde{q}_\sigma$,
$K^R_2=\left(\tilde{k}_{{\bar{\sigma}}[\sigma]}-iZ\right)/\tilde{q}_\sigma$,
for the P [AP] alignment. Here,
${K^{L}_1}^*=(\tilde{k}_{\sigma}-iZ)/\tilde{q}_\sigma$ is the
complex conjugate of $K^L_1$, and $Z={2m\hat{W}}/\hbar ^2
k^{(S)}_F$ is the parameter measuring the strength of each
interface barrier. Approximated wave-vector components, in units
of $k^{(S)}_F$, are $\tilde{q}_\sigma=\sqrt{1-\tilde{\bf
k}^2_{||,\sigma} }$, $\tilde{k}_{\sigma}=\lambda_\sigma
\cos\theta$, and $|\tilde{\bf k}_{||,\sigma}|=\lambda_\sigma
\sin\theta$, where $\lambda_\sigma=\kappa\sqrt{1+\rho_\sigma X}$,
$X=h_0 /E^{(F)}_F\geq 0$, and $\kappa\neq 1$ is measuring FWVM.

In the corresponding FNF double junction, AR and TH processes are
absent, and the expression for NR amplitude, Eq. (\ref{b
general}), reduces to
\begin{equation}
\label{bN} b^N_{\sigma}(E,\theta)=\frac{ ({K^{L}_1}^* -
K^{R}_1)\cos(lq^N_{\sigma})+i (1-{K^{L}_1}^*
K^{R}_1)\sin(lq^N_{\sigma}) }{ (K^{L}_1 +
K^{R}_1)\cos(lq^N_{\sigma})-i(1+K^{L}_1
K^{R}_1)\sin(lq^N_{\sigma}) },
\end{equation}
where $q^N_{\sigma}=\sqrt{(2m/\hbar ^2)(E^{(S)}_F+E)-{\bf
k}^2_{||,\sigma}}$.

To complete our considerations, we also present the probability
amplitudes for an FS single junction in the same notation,
\begin{eqnarray}
\label{a FS} a_\sigma(E,\theta)&=&\frac{2
(\tilde{k}_{\sigma}/\tilde{q}_\sigma) \Delta }{{\cal A}^L_+E +
{\cal B}^L_+\Omega},\\ \label{b FS}
b_\sigma(E,\theta)&=&\frac{{\cal C}_+E + {\cal
D}_+\Omega}{{\cal A}^L_+E + {\cal B}^L_+\Omega},\\
\label{c FS} c_\sigma(E,\theta)&=&\frac{2
(\tilde{k}_{\sigma}/\tilde{q}_\sigma) E\bar{u}(1+K^L_2) }{{\cal
A}^L_+E + {\cal B}^L_+\Omega},\\ \label{d FS}
d_\sigma(E,\theta)&=&\frac{2 (\tilde{k}_{\sigma}/\tilde{q}_\sigma)
E\bar{v}(1-K^L_2)}{{\cal A}^L_+E + {\cal B}^L_+\Omega}.
\end{eqnarray}
Note that $c_\sigma$ and $d_\sigma$ now describe the transmission
of the Bogoliubov electron-like and hole-like quasiparticle,
respectively. The well-known BTK results can be reproduced by
taking $X=0$, $\kappa=1$, and $\theta=0$ in Eqs. (\ref{a
FS})-(\ref{d FS}).\cite{foot}

In case of the simplest NSN metallic junction, taking $X=0$,
$Z=0$, and $\kappa=1$ in Eqs. (\ref{a general})-(\ref{d general}),
the scattering probabilities can be written in an explicit
form\cite{ZZM}
\begin{eqnarray}
\label{aNSN 0}
A_\sigma(E,\theta)&=&\left|\frac{\Delta\sin(\zeta_-/2)}{E\sin(\zeta_-/2)+i\Omega\cos(\zeta_-/2)}\right|^2,\\
\label{bNSN 0} B_\sigma(E,\theta)&=&D_\sigma(E,\theta)=0,\\
\label{cNSN 0}
C_\sigma(E,\theta)&=&\left|\frac{\Omega}{E\sin(\zeta_-/2)+i\Omega\cos(\zeta_-/2)}\right|^2.
\end{eqnarray}

Solutions for the other three types of injection can be obtained
by the same procedure. In particular, if a hole with energy $-E$,
spin $\sigma$, and angle of incidence $\theta$ is injected from
the left, the substitution $q^+_{\sigma} \rightleftharpoons
q^-_{\sigma}$ holds, and the scattering probabilities are the same
as for the injection of an electron with $E$, $\sigma$, and
$\theta$. Therefore, in order to include the description of both
electron and hole injection, the calculated probabilities should
be regarded as even functions of $E$. Also, for an electron or a
hole, injected from the right, the probabilities are the same as
for the injection from the left, except $\sigma\to\bar{\sigma}$
for the AP alignment.

\begin{figure}[b]
\dimen255=0.55\textwidth
    \centerline{\psfig{figure=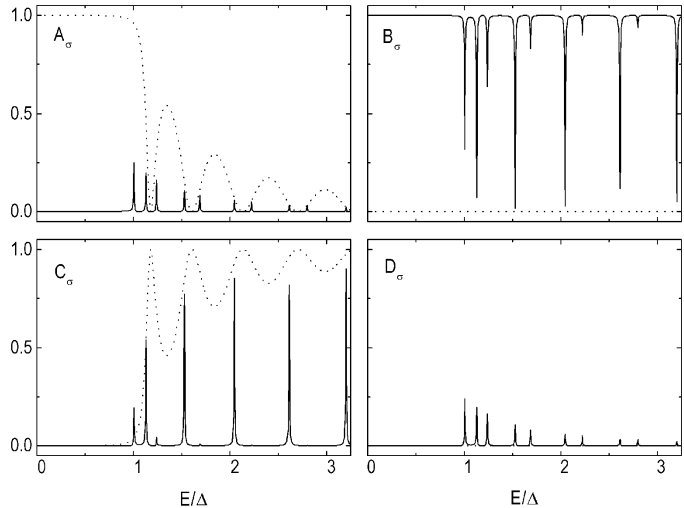,height=0.7\dimen255}}
    \caption{Scattering probabilities, $A_\sigma(E,0)$, $B_\sigma(E,0)$,
            $C_\sigma(E,0)$, and $D_\sigma(E,0)$, for an NSN double junction
            with transparent and highly non-transparent interfaces, $Z=0$
            (dotted curves) and $Z=10$ (solid curves), respectively. The
            parameters are $lk^{(S)}_F=10^4$, $X=0$, $\kappa=1$, and
            $\Delta/E^{(S)}_F=10^{-3}$.}
    \label{tunnel10}
\end{figure}

\begin{figure}[h]
\dimen255=0.55\textwidth
    \centerline{\psfig{figure=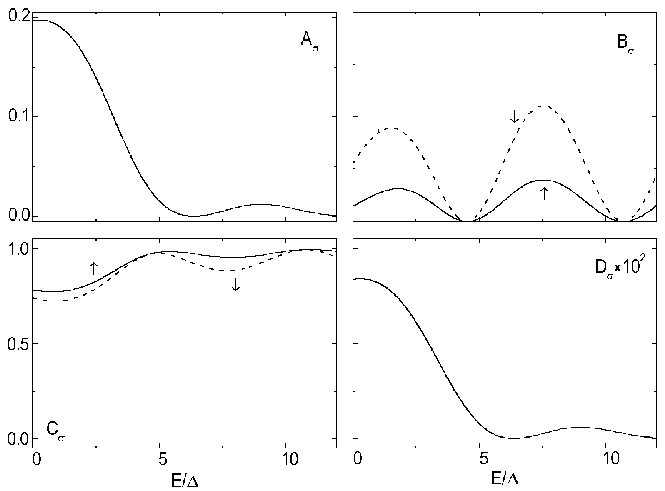,height=0.7\dimen255}}
    \caption{Scattering probabilities, $A_\sigma(E,0)$, $B_\sigma(E,0)$, $C_\sigma(E,0)$,
            and $D_\sigma(E,0)$, for an FSF double junction with thin
            superconducting film, $lk^{(S)}_F=10^3$, for $X=0.5$, $Z=0$,
            $\kappa=1$, $\Delta/E^{(S)}_F=10^{-3}$, and P alignment. Solid
            curves: injection of an electron with $\sigma=\uparrow$. Dashed
            curves: injection of an electron with $\sigma=\downarrow$. }
    \label{thin}
\end{figure}

\begin{figure}[h]
\dimen255=0.55\textwidth
    \centerline{\psfig{figure=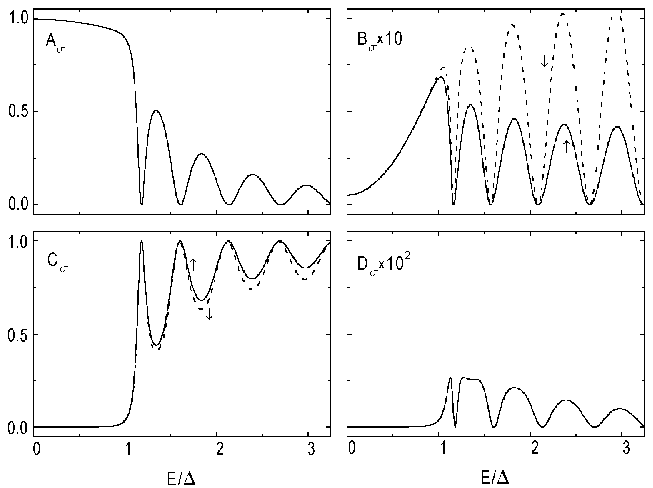,height=0.7\dimen255}}
    \caption{Scattering probabilities, $A_\sigma(E,0)$, $B_\sigma(E,0)$, $C_\sigma(E,0)$,
        and $D_\sigma(E,0)$, for an FSF double junction with thick
        superconducting film, $lk^{(S)}_F=10^4$, for $X=0.5$, $Z=0$,
        $\kappa=1$, $\Delta/E^{(S)}_F=10^{-3}$, and P alignment. Solid
        curves: injection of an electron with $\sigma=\uparrow$. Dashed
        curves: injection of an electron with $\sigma=\downarrow$. }
    \label{thick}
\end{figure}

Following the conservation of ${\bf k}_{||,\sigma}$, transmission
of an electron (hole) with $\sigma=\uparrow$, injected from the
left electrode into the superconductor, is possible only for
angles of incidence $\theta$ satisfying $\theta <\theta _{c1}$,
where $\theta _{c1}=\arcsin(1/\lambda_\uparrow)$ is the angle of
total reflection. Then, $A_\uparrow(E,\theta)=0$ and
$B_\uparrow(E,\theta)=1$ for $\theta >\theta _{c1}$. On the other
hand, $\tilde{k}_{\downarrow}$, which corresponds to the hole
(electron) created by the Andreev reflection, is real only for
$\theta<\theta _{c2}=
\arcsin(\lambda_\downarrow/\lambda_\uparrow)$. The virtual Andreev
reflection occurs for $\theta _{c2}<\theta<\theta _{c1}$, since
$\tilde{k}_{\downarrow}$ becomes imaginary in that case. For
injection of an electron (hole) with $\sigma=\downarrow$,
transmission into the superconductor is possible for any
$\theta<\pi/2$, and $\tilde{k}_{\uparrow}$ is always
real.\cite{Beasley}

From Eqs. (\ref{a general}) and (\ref{d general}) it follows that
$A_\sigma(E,\theta)=D_\sigma(E,\theta)=0$ when
\begin{equation}
\label{resonance} \zeta_-=2n\pi
\end{equation}
for $n=0,\pm 1,\pm 2,\ldots$. Therefore, the Andreev reflection at
both interfaces vanishes at the energies of geometrical resonances
in quasiparticle spectrum. The effect is similar to the
over-the-barrier resonances in the simple problem of one-particle
scattering against a step-function potential,\cite{CohenT} the
superconducting gap playing the role of a finite-width barrier (as
in the semiconductor model\cite{Tinkham}). The absence of AR and
TH processes means that all quasiparticles with energies
satisfying Eq. (\ref{resonance}) will pass unaffected from one
electrode to another, without creation or annihilation of Cooper
pairs.

Characteristic features of coherent quantum transport through
clean superconducting layers are the subgap tunneling and
oscillations of the scattering probabilities. For $E<\Delta$, the
subgap tunneling suppresses the Andreev reflection, thereby
enhancing the transmission. For $E>\Delta$, all probabilities
oscillate with $E$ and $l$ due to the interference effect.

The interface resistance reduces AR and TE, and enhances NR and TH
probabilities. In contrast to the positions of zeros of
$A_\sigma(E,0)$, given by Eq. (\ref{resonance}), the positions of
maxima of $A_\sigma(E,0)$, as well as that of zeros and maxima of
$B_\sigma(E,0)$, $C_\sigma(E,0)$, and $D_\sigma(E,0)$, are
$Z$-dependent. Approaching the tunnel limit ($Z\to\infty$), peaks
in the scattering probabilities gradually split into two spikes
belonging to consecutive pairs with positions defined by the
quantization conditions
\begin{equation}
\label{n} lq^+_{\sigma}=n_1\pi,~~~lq^-_{\sigma}=n_2\pi.
\end{equation}
Here, $n_1-n_2=2n$, with $n$ coresponding to that of Eq.
(\ref{resonance}). The exception is the spike at the gap edge,
originating from the singularity in the BCS density of states.
Note that Eq. (\ref{n}) gives the bound state energies of an
isolated superconducting film. This is illustrated in Fig.
\ref{tunnel10} for an NSN junction, showing a simple connection
between the resonances in metallic junctions ($Z=0$) and the bound
states in the corresponding tunnel junctions ($Z\to\infty$).

The influence of exchange interaction is illustrated in Figs.
\ref{thin} and \ref{thick} for an FSF double junction in P
alignment with $Z=0$ and $\kappa=1$. Taking
$\Delta/E^{(S)}_F=10^{-3}$, in a thin superconducting film,
$lk^{(S)}_F\sim 10^3$, the Andreev reflection is strongly
suppressed, since the subgap transmission is considerable, Fig.
{\ref{thin}}. In this case, the oscillations are less pronounced,
with the period much larger than $\Delta$. For a thick film,
$lk^{(S)}_F\sim 10^4$, the subgap tunneling is irrelevant (except
for small 'tails' in $A_\sigma(E,0)$ and $C_\sigma(E,0)$ at
$E\lesssim\Delta$) and above the gap the oscillations are
pronounced, with the period on the order of $\Delta$, Fig.
{\ref{thick}}. The scattering probabilities for AP alignment
differ very slightly in the case of normal incidence, $\theta=0$.
Although spin-independent due to the singlet-state pairing,
$A_\sigma(E,0)$ is suppressed in comparison with the corresponding
NSN junction, and $D_\sigma(E,0)$ becomes non-trivial. The
spin-dependent normal reflection also occurs, $B_\sigma(E,0)$
having zeros at the same energies as $A_\sigma(E,0)$ and
$D_\sigma(E,0)$, so that maxima in $C_\sigma(E,0)$ are still equal
to unity due to the interface transparency.

\bigskip

\section{Differential conductances}

When voltage $V$ is applied to the junction, the charge current
density is given by
\begin{equation}
\label{j} j_q(V)=\sum_{\sigma}\int\frac{d^3{\bf k}}{(2\pi)^3}
e{\bf v}\cdot {\bf \hat{z}}~\delta f({\bf k},V),
\end{equation}
where ${\bf v}=(\hbar/m)\Im[u_\sigma^*({\bf r})\nabla
u_\sigma({\bf r})+v_{\bar{\sigma}}^*({\bf r})\nabla
v_{\bar{\sigma}}({\bf r})]$ is the velocity, and $\delta f({\bf
k},V)$ is the asymmetric part of the nonequilibrium distribution
function of current carriers. Using the solution of the scattering
problem for the injection of an electron from the left, described
in the previous section, and the dispersion relation ${\bf
k}(E)=k^{+}_{\sigma}{\bf \hat{z}}+{\bf k}_{||,\sigma}$, Eq.
(\ref{j}) can be rewritten in the form
\begin{equation}
j_q(V)=\frac{e{k^{(S)}_{F}}^2}{\pi
h}\int\limits_{-\infty}^{\infty}dE \sum_{\sigma}\lambda^2_\sigma
\int\limits_{0}^{\pi/2}d\theta \sin\theta \cos\theta
\left[1+A_\sigma(E,\theta)-B_\sigma(E,\theta)\right]\delta f({\bf
k},V).
\end{equation}
In accordance with BTK, without solving the suitable transport
equation, we take $\delta f({\bf k},V)=f_0(E-eV/2)-f_0(E+eV/2)$,
where $f_0(E)$ is the Fermi-Dirac equilibrium distribution
function.

\begin{figure}[b]
\dimen255=0.7\textwidth
    \centerline{\psfig{figure=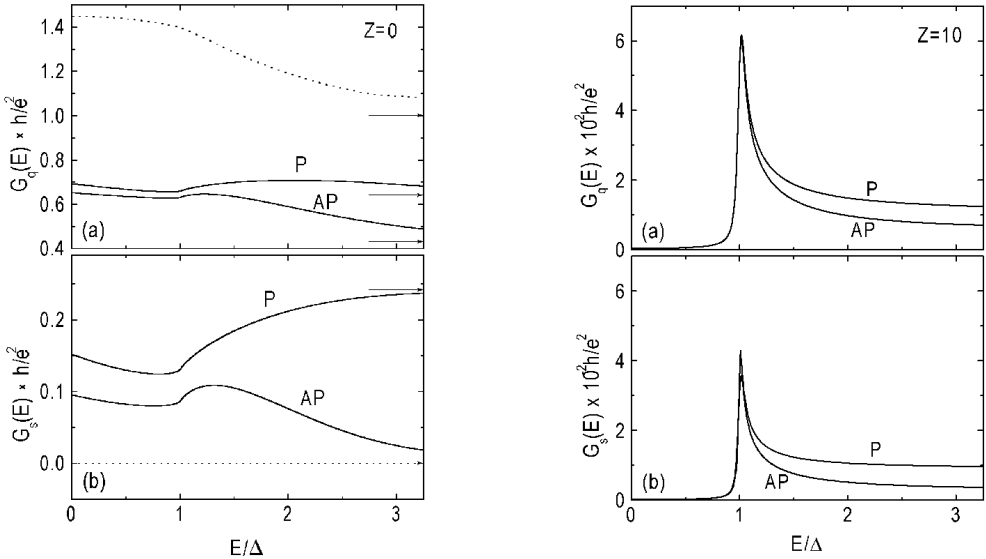,height=0.7\dimen255}}
    \caption{{\bf (Left panel)} Differential charge (a) and spin
            (b) conductance spectra, $G_q(E)$ and $G_s(E)$, of an FSF double
            planar junction with thin superconducting film, $lk^{(S)}_F=10^3$,
            for $X=0.5$, $Z=0$, $\kappa=1$, $\Delta/E^{(S)}_F=10^{-3}$, in P
            and AP alignment. Conductances of the corresponding NSN junction
            are shown for comparison (dotted curves). Arrows indicate $G^N_q$
            and $G^N_s$ values.  }
    \label{l3}
    \caption{{\bf (Right panel)} Differential charge (a) and spin
            (b) conductance spectra, $G_q(E)$ and $G_s(E)$, of an FSF double
            tunnel junction with thin superconducting film, $lk^{(S)}_F=10^3$,
            for $X=0.5$, $Z=10$, $\kappa=1$, $\Delta/E^{(S)}_F=10^{-3}$, in P
            and AP alignment. }
    \label{Z10 thin}
\end{figure}

\begin{figure}[htb]
\dimen255=0.7\textwidth
    \centerline{\psfig{figure=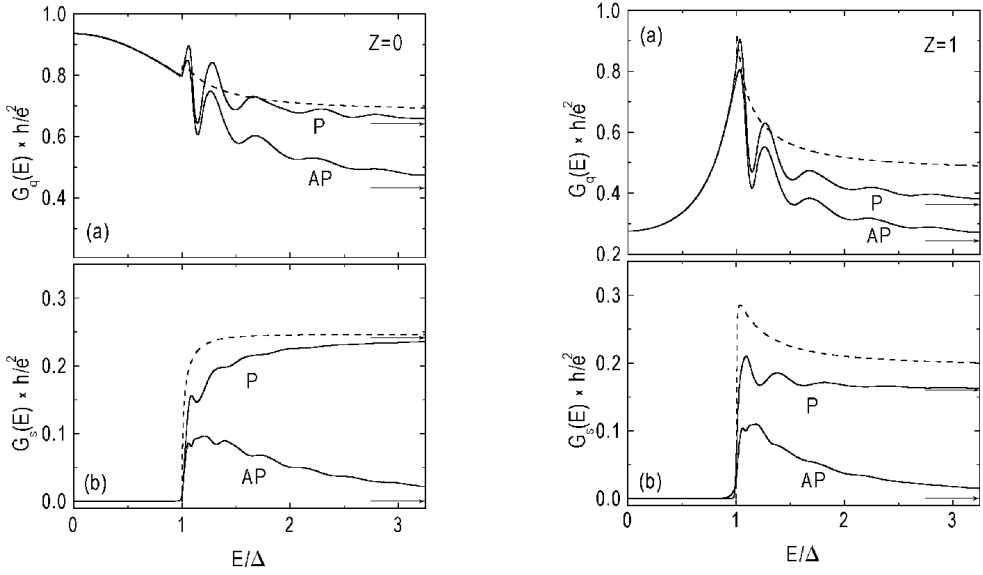,height=0.7\dimen255}}
    \caption{{\bf (Left panel)} Differential charge (a) and spin
            (b) conductance spectra, $G_q(E)$ and $G_s(E)$, of an FSF double
            planar junction with thick superconducting film,
            $lk^{(S)}_F=10^4$, for $X=0.5$, $Z=0$, $\kappa=1$,
            $\Delta/E^{(S)}_F=10^{-3}$, in P and AP alignment. Dashed curves
            represent the generalized BTK results for the same parameters.
            Arrows indicate $G^N_q$ and $G^N_s$ values. }
    \label{GZ0}
    \caption{{\bf (Right panel)} Differential charge (a) and spin
            (b) conductance spectra, $G_q(E)$ and $G_s(E)$, of an FSF double
            planar junction with thick superconducting film,
            $lk^{(S)}_F=10^4$, for $X=0.5$, $Z=1$, $\kappa=1$,
            $\Delta/E^{(S)}_F=10^{-3}$, in P and AP alignment. Dashed curves
            represent the generalized BTK results for the same parameters.
            Arrows indicate $G^N_q$ and $G^N_s$ values. }
    \label{GZ1}
\end{figure}

In this approach, the charge current per electron is given by
\begin{equation}
\label{Iq} I_q(V)=\frac{1}{e}\int\limits_{-\infty}^{\infty}dE
\left[f_0(E-eV/2)-f_0(E+eV/2)\right]G_q(E),
\end{equation}
where the spin-averaged differential charge conductance at zero
temperature is
\begin{equation}
\label{3D q} G_q(E)=\frac{e^2}{2h}\sum_\sigma \lambda^2_\sigma
\int\limits_{0}^{\pi/2}d\theta~\sin\theta\cos\theta
\left[1+A_\sigma(E,\theta)-B_\sigma(E,\theta)\right].
\end{equation}
By analogy, the corresponding spin current (proportional to the
probability current) is given by
\begin{equation}
\label{Is}
I_s(V)=\frac{1}{e}\int\limits_{-\infty}^{\infty}dE\left[f_0(E-eV/2)-f_0(E+eV/2)\right]G_s
(E),
\end{equation}
where the differential spin conductance at zero temperature is
\begin{equation}
\label{3D s} G_s(E)=\frac{e^2}{2h}\sum_\sigma\rho_\sigma
\lambda^2_\sigma
\int\limits_{0}^{\pi/2}d\theta~\sin\theta\cos\theta
\left[1-A_\sigma(E,\theta)-B_\sigma(E,\theta)\right].
\end{equation}
Note that in Eqs. (\ref{3D q}) and (\ref{3D s}) the upper limit of
integration over $\theta$ for $\sigma=\uparrow$ is the angle of
total reflection $\theta_{c1}\leq\pi/2$. Avoiding integration over
$\theta$, the differential charge and spin conductances of a
point-contact FSF double junction are simply expressed by
\begin{equation}
\label{PC q} G_q(E)=\frac{e^2}{2h}\sum_\sigma \lambda^2_\sigma
\left[1+A_\sigma(E,0)-B_\sigma(E,0)\right]
\end{equation}
and
\begin{equation}
\label{PC s} G_s(E)=\frac{e^2}{2h}\sum_\sigma\rho_\sigma
\lambda^2_\sigma \left[1-A_\sigma(E,0)-B_\sigma(E,0)\right].
\end{equation}

The influence of the exchange interaction on the conductance
spectra is shown for $X=0.5$ on the example of thin,
$lk^{(S)}_F=10^3$, and thick, $lk^{(S)}_F=10^4$, superconducting
films (Figs. \ref{l3}-\ref{GZ1}). Besides the case of transparent
interfaces, $Z=0$ (Figs. \ref{l3} and \ref{GZ0}), the effect of
interface resistance is illustrated in the tunnel limit, $Z=10$,
and for weak non-transparency, $Z=1$, in Figs. \ref{Z10 thin} and
\ref{GZ1}. The influence of FWVM on the conductance spectra,
$\kappa\neq 1$, is similar to that of the interface
resistance.\cite{Milos} The values of normal conductances, $G^N_q$
and $G^N_s$ of the corresponding FNF double planar junction,
indicated by arrows, are obtained by setting
$A_\sigma(E,\theta)=0$ and
$B_{\sigma}(E,\theta)=\left|b^N_{\sigma}(E,\theta)\right|^2$ in
Eqs. (\ref{3D q}) and (\ref{3D s}), where $b^N_{\sigma}(E,\theta)$
is given by Eq. (\ref{bN}).

\begin{figure}[t]
\dimen255=0.55\textwidth
    \centerline{\psfig{figure=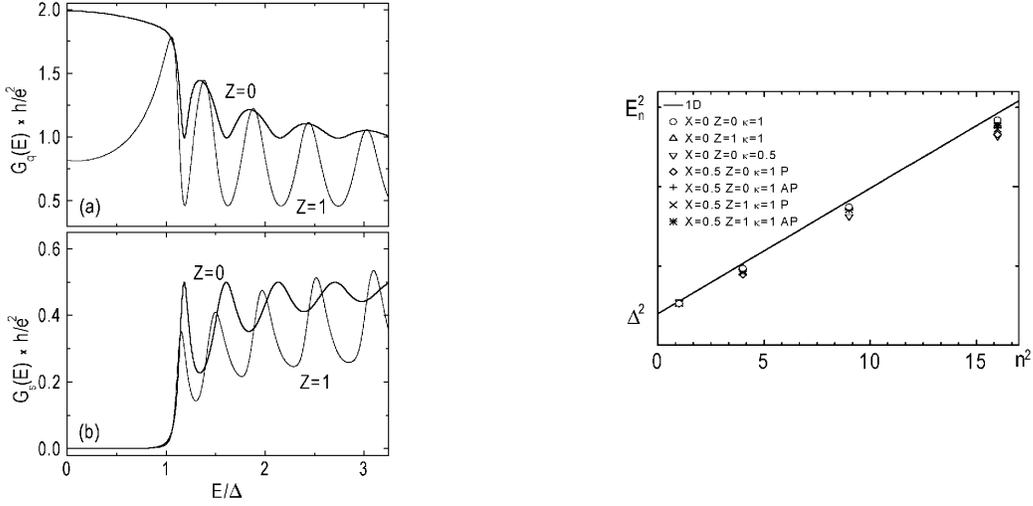,height=0.7\dimen255}}
    \caption{{\bf (Left panel)} Differential charge (a) and spin
            (b) conductance spectra, $G_q(E)$ and $G_s(E)$, of a point-contact
            FSF double junction, with P alignment, for the same parameters as
            in Figs. 7 and 8 ($X=0.5$, $\Delta/E^{(S)}_F=10^{-3}$,
            $lk^{(S)}_F=10^4$, and $\kappa=1$). Here, the difference between
            conductances for the AP and the P alignment is negligible. }
    \label{1D}
    \caption{{\bf (Right panel)} Square of resonant
            energies $E^2_n$ as a function of $n^2$, obtained from positions
            of minima of $G(E)$ for $lk_F^{(S)}=10^4$. The intercept is
            $\Delta^2$ and the slope is $(hv^{(S)}_F/2l)^2$. }
    \label{E2n2}
\end{figure}

The spin-polarized subgap tunneling of quasiparticles, and strong
suppression of the Andreev reflection as a consequence, is
significant in thin superconducting films, whereas the conductance
oscillations above the gap are pronounced in the thick films. The
magnetoresistance is apparent, as charge and spin conductances are
larger for the P than for the AP alignment. An important
consequence of the coherent transport is that the spin conductance
is non-trivial for the AP alignment, approaching its normal value
$G^N_s=0$ either for $E/\Delta\gg 1$, or in the tunnel limit
($Z\to\infty$) for all energies. We emphasize that the amplitudes
of the oscillations are considerably larger for the point-contact
FSF than for the planar FSF double junction, Fig. \ref{1D}.

Incoherent transport through an FSF double junction is described
as a transport through the corresponding FS and SF junctions in
series. In that case, the conductance spectra are calculated using
the generalized BTK probabilities, obtained from Eqs. (\ref{a FS})
and (\ref{b FS}). Numerical results for the incoherent transport
are also presented in Figs. \ref{GZ0} and \ref{GZ1} for
comparison. It is evident that in thick films the only difference
comes from the interference-effect oscillations for the energies
above the gap. In contrast with the coherent transport, for the AP
alignment $G_s(E)\equiv 0$, and nonequilibrium spin density
accumulation changes the chemical potential of two spin subbands
in the superconductor. This reduces the superconducting gap with
increasing voltage, and destroys the superconductivity at a
critical voltage on the order of $\Delta/e$.\cite{Takahashi} The
effect of incoherency is less pronounced in metallic than in the
tunnel junctions due to the Andreev reflection.\cite{Kinezi}

The results can be applied to reliable spectroscopic measurements
of $\Delta$ and $v^{(S)}_F$ in superconducting films. From Eq.
(\ref{resonance}), for $\theta=0$, the energy $E_n$ of the $n$-th
geometrical resonance (conductance minimum) satisfies a simple
relation
\begin{eqnarray}
\label{En} E_n^2=\Delta^2+\Big(\frac{hv^{(S)}_F}{2l}\Big)^2n^2.
\end{eqnarray}
Therefore, the linear plot of $E_n^2$ vs $n^2$ has the intercept
equal to $\Delta^2$ and the slope equal to $(hv^{(S)}_F/2l)^2$. An
example is shown in Fig. \ref{E2n2}. Note that even the points
obtained for planar (3D) double junctions lie almost on the same
straight line given by Eq. (\ref{En}) for the particular case of a
point-contact (1D) double junction. The numerical results show
that the method is almost independent on dimensionality of the
junction and on parameters of the electrodes.

\begin{figure}[htb]
\dimen255=0.7\textwidth
    \centerline{\psfig{figure=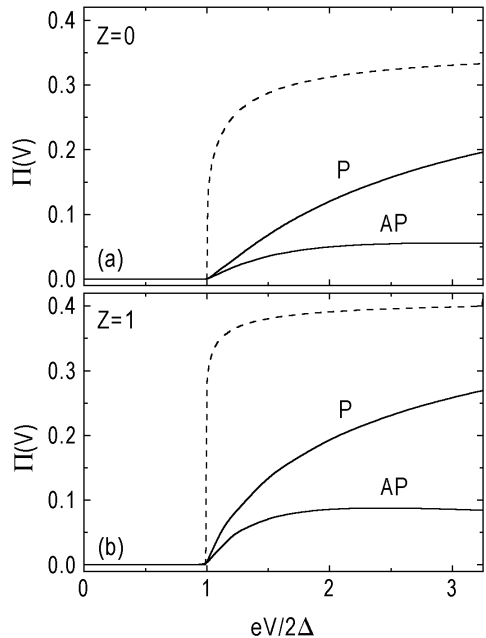,height=0.8\dimen255}}
    \caption{Spin polarization of the current, $\Pi(V)$, for an FSF double planar
            junction with P and AP alignment, is shown for (a) $Z=0$, and (b)
            $Z=1$. Other parameters are the same as in Figs. \ref{GZ0} and
            \ref{GZ1} ($X=0.5$, $\Delta/E^{(S)}_F=10^{-3}$, $lk^{(S)}_F=10^4$,
            and $\kappa=1$). Dashed curves represent the generalized BTK
            results.}
    \label{Pi}
\end{figure}

The net spin polarization of the current is defined as $\Pi
(V)={I_s (V)}/{I_q (V)}$. In thin superconducting films, $\Pi (V)$
is almost constant, considerably smaller than $X$, which is the
polarization in the corresponding FNF junction. Below the gap,
$eV/2\Delta<1$, the contribution of subgap tunneling to spin
polarization becomes negligibly small for large $l$. Above the
gap, $eV/2\Delta>1$, the polarization increases with $V$, the
increase becoming steeper with the interface non-transparency,
Fig. \ref{Pi}. On the other hand, in a tunnel FS junction the
polarization abruptly changes from $\Pi=0$ to $\Pi=X$ at
$eV/2\Delta=1$. The same result holds for incoherent transport
through an FSF double tunnel junction.

\bigskip

\section{Summary}

We have analyzed transport properties of FSF double-barrier
junctions, taking into account the influence of the exchange
interaction, the resistance of the interfaces, and the Fermi
velocity mismatch on the scattering probabilities and the
conductance spectra. We have shown that subgap tunneling and
oscillations of differential conductances are the main features of
the coherent quantum transport through a superconducting layer in
both FSF and NSN double-barrier junctions. The subgap tunneling
suppresses the Andreev reflection, thereby enhancing the
transmission, especially in thin films. The scattering
probabilities and conductances oscillate as a function of the
layer thickness and of the quasiparticle energy above the gap.

Periodic vanishing of the Andreev reflection at the energies of
geometrical resonances is found as an important consequence of the
quasiparticle interference. Insulating barriers at the interfaces
reduce the Andreev reflection and transmission, mainly for
energies below the gap. The Fermi velocity mismatch has a similar
effect. Results are directly accessible to experiments. In
principle, oscillations of differential conductances with the
period of geometrical resonances could be used for reliable
spectroscopy of quasiparticle excitations in superconductors.

In conclusion, finite-size effects, along with the difference
between coherent and incoherent transport, are essential for
spin-currents in FSF junctions. For the coherent transport,
besides the spin-polarized subgap tunneling in thin
superconducting films, pronounced oscillations of spin conductance
occur in thick films. As a consequence of the quasiparticle
interference, a non-trivial spin current without the excess spin
accumulation and destruction of superconductivity is possible even
for AP alignment of the electrode magnetizations.

\bigskip

\section{Acknowledgment}

We are grateful to Ivan Bo\v{z}ovi\'c for pointing out the
significance of the problem treated in this paper, and for help at
the initial stage of this work. Furthermore, we thank Irena
Kne\v{z}evi\'c for useful discussions. This work has been
supported by the Serbian Ministry of Science, Technology, and
Development, grant ${\rm N}^{\circ} 1899$.



\begin{thebibliography}{99}
\bibitem{foot} The parameter $Z$ in our definition
differs from the BTK by a factor of 2.
\end{thebibliography}
\end{document}